\documentclass[aps,prb,onecolumn,nofootinbib,citeautoscript,10pt, showkeys]{revtex4-2}

\usepackage{amsmath,amssymb} 
\usepackage{comment}

\usepackage[tight]{subfigure} 

\usepackage[dvipsnames]{xcolor} 
\usepackage[papersize={8.5in,11in}]{geometry}
\usepackage[colorlinks=true]{hyperref}
\hypersetup{
    bookmarks=true,         
    unicode=false,          
    pdftoolbar=true,        
    pdfmenubar=true,        
    pdffitwindow=false,     
    pdfstartview={FitH},    
    pdfkeywords={keyword1} {key2} {key3}, 
    pdfnewwindow=true,      
    colorlinks=true,       
    linkcolor=magenta, 
    citecolor=blue,        
    filecolor=magenta,      
    urlcolor=blue           
} 

\usepackage{dcolumn}
\usepackage{color}
\usepackage{amssymb,amsmath}
\usepackage{tabularx,graphicx}
\usepackage{epstopdf}
\usepackage{latexsym}
\usepackage{colortbl}
\usepackage{psfrag}
\usepackage{bbm,bm,array,physics}
\usepackage{dsfont}
\usepackage{float, mathrsfs}

\def \nn{\nonumber \\}
\def\*#1{\boldsymbol{#1}} 

\begin{document}

\title{Distinguishing features of longitudinal magnetoconductivity for a Rarita-Schwinger-Weyl node}

\author{Ipsita Mandal}
\email{ipsita.mandal@snu.edu.in}

\affiliation{Department of Physics, Shiv Nadar Institution of Eminence (SNIoE), Gautam Buddha Nagar, Uttar Pradesh 201314, India}

\begin{abstract} 
The band-degeneracy points in the Brillouin zones of chiral crystals exist in multiple avatars, with the high-symmetry points being able to host multifold nodes of distinct characters. A class of such crystals, assisted by the spin-orbit coupling, harbours fourfold degeneracy in the form of Rarita-Schwinger-Weyl node (RSWN) at the $\Gamma$-point. Our aim is to explore the nature of longitudinal magnetoconductivity, arising from applying collinear electric and magnetic fields, for such systems. Adjusting the chemical potential to lie near the intrinsic energy-location of the RSWN, the multifold nature of the RSWN is revealed by an interplay of intraband and interband scatterings, which would not arise in twofold degeneracies like the conventional Weyl nodes. The current study fills up the much-needed gap in obtaining the linear response from an exact computation, rather than the insufficient relaxation-time approximation employed earlier. 
\end{abstract}


\maketitle

\tableofcontents

\section{Introduction}

The unprecedented advancements on the experimental fronts, coupled with the DFT studies of materials, have put forward numerous generalisations of the archetypal Weyl semimetal (WSM) of the three-dimensional (3d) variety\cite{burkov11_Weyl, armitage_review, yan17_topological}. The excitement surrounding these so-called semimetallic systems mainly stems from the fact that their Brillouin zones (BZs) harbour nodal points, which, in turn, furnish the momentum space with the mathematical properties of topology \cite{berry, xiao_review, sundaram99_wavepacket, graf-Nband}, when viewed as closed manifolds. Physical properties like the Berry curvature (BC) and the orbital magnetic moment (OMM) quantify the topological properties lurking beneath, which are detectable in transport experiments \cite{ips_rahul_ph_strain, rahul-jpcm, ips-ruiz, ips-floquet,ips-tilted, claudia-multifold, timm,  li_nmr17, ips-spin1-ph, ips-exact-spin1, ips-nl-ph}. Generalising the simplest prototype of an isotropic and linear-in-momentum twofold band-crossing, a pseudospin-3/2 Rarita-Schwinger-Weyl node (RWSN)~\cite{bernevig, long, igor, igor2, isobe-fu, tang2017_multiple, ips3by2, ips-jns, ips-cd1,ips_cd, ma2021_observation, ips-magnus, ips_jj_rsw} represents a fourfold band-crossing at a high-symmetry point of the BZ of a chiral crystal \cite{bernevig, chang2018, grushin-multifold}. Adding to the cornucopia of transport properties that can reflect the exquisite character of RSWNs \cite{ips-magnus, ips-cd1, ips-jns, ips_jj_rsw, ips-rsw-ph, ips-shreya, ips-internode}, we investigate here the longitudinal magnetoconductivity arising from an isolated node \cite{ips-exact-kwn}.

The perks of invoking the analogy of the BC (existing in the momentum space) to a magnetic field (existing in the physical or real space) is that we conjure up the intuitive picture of nonzero BC monopoles sitting at the nodal points \cite{fuchs-review, polash-review}, acting as the sources and sinks of the vector field of the BC-flux. Utilising the mathematical machinery of algebraic topology, the monopoles' charges also represent the so-called Chern numbers about the nodal points. The sign of $\mathcal C$ is pictured as the chirality ($\chi$) of the node, conferring a notion of \textit{handedness or chirality} on the quasiparticles populating the bands converging there. Appropriately, they are termed as \textit{right-handed} or \textit{left-handed} quasiparticles, depending on whether $\chi = 1$ or $\chi = -1$.

The appearance of chirally-charged nodes in 3d crystal-lattices is governed by the Nielsen-Ninomiya theorem \cite{nielsen81_no}, which is physically reflected by the fact that the sum of the Chern numbers over the entire BZ must yield zero. For nodes arising in achiral crystals (e.g., TaAs family \cite{lv_Weyl}), such conjugate partners are typically (almost) degenerate in energy, due to the presence of mirror or other roto-inversion symmetries. Therein, charge-pumping is an important internode phenomenon in the presence of collinear external electric ($\boldsymbol E $) and magnetic ($\boldsymbol B $) fields. The phenomenon in question is an embodiment of the chiral anomaly in the arena of condensed matter physics \cite{chiral_ABJ, son13_chiral, li_nmr17, ips-internode}, where the analog of the well-known concept of spacetime chirality (of relativistic fermions) is realised in the momentum space. It provides a sharp contrast with the oppositely-charged chiral nodes in chiral crystals, where they need not be degenerate in energy, because the conjugate nodes are not related by crystal symmetries. In fact, they are observed to have discernible separations in energy and momenta, with isolated nodes sitting at their intrinsic chemical potentials \cite{law-ksm, chang2018, multifold-rhsi}. The inevitable outcome is that the internode-scattering-induced charge-pumping gets irrelevant. However, at the same time, the energy-offset between the nodes of opposite chiralities fulfills the requisite conditions for observing other important phenomena like quantised circular photogalvanic effect \cite{bernevig, chang2018, ni2021_giant, moore18_optical, guo23_light,kozii, ips_cpge} and circular dichroism \cite{ips-cd1,ips_cd}.

\begin{figure*}[t]
\subfigure[]{\includegraphics[width = 0.25 \textwidth]{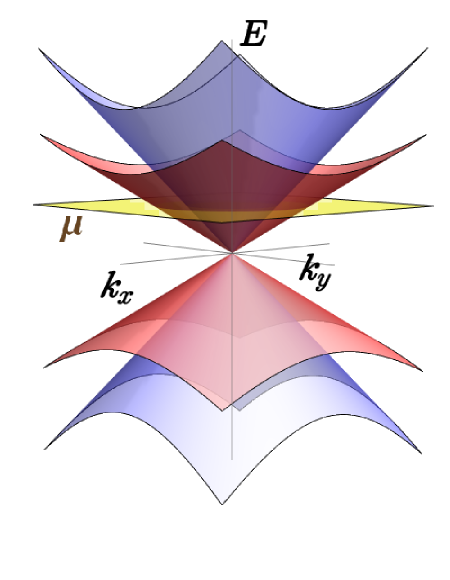}} \hspace{3 cm}
\subfigure[]{\includegraphics[width = 0.32 \textwidth]{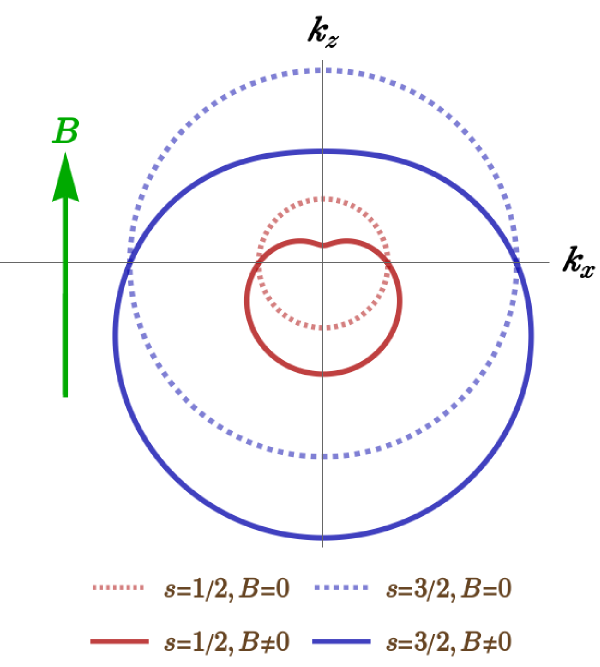}}
\caption{\label{figdis}Schematics of the energy bands meeting at an isotropic RSWN: (a) Bare dispersion ($E$) for $s= \pm 1/2$ (light red) and $s= \pm 3/2$ (light blue) bands against the $k_x k_y$-plane (or, equivalently, $k_y k_z$- and $k_z k_x$-planes). The yellow plane depicts a positive chemical potential ($\mu$) cutting the $s=1/2$ and $s=3/2$ bands, giving rise to two concentric Fermi surfaces. (b) Fermi-surface projections in the $k_z k_x $-plane, when $\boldsymbol B$ (green arrow) is applied along the $z$-axis. To provide an eye-estimate, the dotted curves show the unperturbed projections in the absence of an external magnetic field.}
\end{figure*}

In our recent works, we have addressed the pertinent problem of computing longitudinal magnetoconductivity in systems where multiple bands contribute to transport at a particular node, encompassing the cases of chirally-conjugate nodes
\cite{ips-exact-spin1} and isolated Kramers-Weyl nodes \cite{ips-exact-kwn} (the latter arising exclusively in chiral crystals \cite{hasan-review, long-ksm}). In this paper, we take up a similar problem where an isotropic RSWN (sitting at the $\Gamma$-point) is the system under consideration. As pointed out earlier, only the isolated RWSN will be relevant here for computing the linear response, since we will consider tuning the external chemical potential to lie near the intrinsic energy of the concerned nodal point, safely assuming that the chirally-conjugate node is sufficiently separated in energy and momenta. For example, in RhSi, there exist fourfold- and sixfold-degenerate nodes at the $\Gamma$- and $R$-points of the BZ, carrying Chern numbers of $\pm 4$, and featuring the largest possible momentum-separation between the nodal-points' locations \cite{sanchez, multifold-rhsi}. While the former represents an RSWN, the bands at the latter are doubly-degenerate, featuring a pair of pseudospin-1 triple-point fermions (TPFs). The energy-offset between these nodes is $\sim 0.4 $ eV. All these make it possible to observe really long Fermi arcs, spanning the distance constrained by the separation of the multiply-degenerate bulk-nodes, and topologically protected by their bulk chiral charges.
Other examples of chiral crystals include AlPt \cite{schroter}, RhGe (with the energy-separation between RSWN and the TPFs being $\sim 0.21 $ eV) \cite{bahadur-RhGe}, RhSi \cite{sanchez} Li$_2$Pd$_3$B \cite{law-LiPdB}, Li$_2$Pt$_3$B \cite{law-LiPdB}, and CoSi \cite{sanchez, tang2017_multiple}.

A typical RSWN is depicted schematically in Fig.~\ref{figdis}, with the four bands visualised near the degeneracy point. Exploiting the universal applicability of the Boltzmann equations as the kinetic theory for transport \cite{mermin, ips-kush-review}, we will carry out the exercise of computing the longitudinal conductivity exactly, arising from applying collinear electromagnetic fields ($\boldsymbol   E  \parallel \boldsymbol B $). The framework will be limited to the realm of weak (implying nonquantising) magnetic fields, so that the splitting of the dispersion into discrete Landau-levels can be ignored. Improving on our earlier ventures along similar lines~\cite{ips-rsw-ph, ips-internode}, we will go beyond the relaxation-time approximation (RTA), thereby solving the semimeclassical Boltzmann equations exactly. To elaborate on the context further, we would like to point out that the nature of in-plane and out-of-plane components of conductivity in planar-Hall set-ups have been derived in Refs.~\cite{ips-rsw-ph, ips-internode}, within the simplifying framework of the relaxation-time approximation, considered the aspects of intranode \cite{ips-rsw-ph} and internode \cite{ips-internode} scatterings. The current calculations will help us fix the shortcomings arising from the assumption of a momentum-independent relaxation-time.

The paper is organised as follows: In Sec.~\ref{secmodel}, we describe the explicit form of the low-energy effective Hamiltonian in the vicinity of an RSWN node and the emerging topological properties. Sec.~\ref{secboltz} is devoted to deriving the equations leading to the final values of the longitudinal magnetoconductivity. The results are discussed in Sec.~\ref{secres} therein, illustrated by representative plots. In Sec.~\ref{seccomp}, we discuss our results derived here with those obtained using RTA. 
Finally, we end with a summary and some future perspectives in Sec.~\ref{secsum}. In all our expressions, we will adopt the natural units, which involves setting the reduced Planck's constant ($\hbar $), the speed of light ($c$), and the Boltzmann constant ($k_B $) to unity. Additionally, the electric charge has no units. Despite the fact that the magnitude ($e$) of a single unit of electrons' charge is unity, we will retain it in our expressions solely for the purpose of book-keeping.

\section{Model}
\label{secmodel}

Applying the widely-used method of linearising the $\boldsymbol{k} \cdot \boldsymbol {p}$ Hamiltonian to the fourfold degeneracy at the $\Gamma$-point of a chiral crystal, one obtains the generic forms of the RWSNs \cite{multifold-rhsi, xu2020_optical}. Adopting the isotropic version for the sake of simplicity, we use the model Hamiltonian captured by
\begin{align}
	\mathcal{H}(\boldsymbol{k}) =  v_F \left( k_x\,	{\mathcal J }_x + k_y\, {\mathcal J }_y
+  k_z \, {\mathcal J }_z \right),
\end{align}
where $ \boldsymbol{\mathcal J } = \lbrace {\mathcal J }_x,\, {\mathcal J }_y,\, {\mathcal J }_z \rbrace $ represents the vector operator whose three components comprise the angular-momentum operators in the spin-$3/2$ representation of the SU(2) group. We choose the commonly-used representation with
\begin{align}
{\mathcal J }_x= 
\begin{pmatrix}
	0 & \frac{\sqrt{3}}{2} & 0 & 0 \\
	\frac{\sqrt{3}}{2} & 0 & 1 & 0 \\
	0 & 1 & 0 & \frac{\sqrt{3}}{2} \\
	0 & 0 & \frac{\sqrt{3}}{2} & 0 
\end{pmatrix} , \quad
{\mathcal J }_y=
\begin{pmatrix}
	0 & \frac{-i \,  \sqrt{3}}{2}  & 0 & 0 \\
	\frac{i \, \sqrt{3}}{2} & 0 & -i & 0 \\
	0 & i & 0 & \frac{-i \, \sqrt{3}}{2}  \\
	0 & 0 & \frac{i \, \sqrt{3}}{2} & 0 
\end{pmatrix}, \quad
{\mathcal J }_z =
\begin{pmatrix}
	\frac{3}{2} & 0 & 0 & 0 \\
	0 & \frac{1}{2} & 0 & 0 \\
	0 & 0 & -\frac{1}{2} & 0 \\
	0 & 0 & 0 & -\frac{3}{2} 
\end{pmatrix}.
\end{align}

The energy eigenvalues are found to be
\begin{align}
\label{eqeval}
\varepsilon_s ( k ) =  s \, v_F \, k \,, \quad 
s \in  \left \lbrace  \pm \frac{1}{2}, \pm \frac{3}{2} \right \rbrace,
\end{align}
where $ k = \sqrt{k_x^2 + k_y^2 + k_z^2 } $.
Thus, each of the four bands demonstrates an isotropic linear-in-momentum dispersion  like a Weyl cone [cf. Fig.~\ref{figdis}(a)]. The signs of ``$+$'' and ``$-$'' give us the dispersion relations for the conduction and valence bands, respectively, with respect to the nodal point. The corresponding group-velocities of the quasiparticles are given by 
\begin{align} 
\boldsymbol{v}_{s}(\boldsymbol{k}) &= 
\nabla_{\boldsymbol{k}}  \varepsilon_{s}(\boldsymbol{k})  
=  \frac{ s \, v_F \,\boldsymbol k }{k} \, .  
\end{align}
The isotropy of $\mathcal H $ indicates that the required calculations can be simplified by adopting the spherical-polar coordinates. Hence, we will be the employing the transformations involving
\begin{align}
\label{eqcyln} 
k_x =  k \sin \theta  \cos \phi \,, \quad
k_y =   k \sin \theta  \sin \phi \,, \quad k_z = k \cos \theta\,,
\end{align}
where $ k \in [0, \infty )$, $\phi \in [0, 2 \pi )$, and $\theta \in [0, \pi ]$. 

\subsection{Eigenvectors}

For an orthonormal set of eigenvectors, $ \lbrace \psi_{s} (\boldsymbol k) \rbrace $, the explicit forms of the positive-energy bands are given by
\begin{align}
\label{eqev}
 \psi_{1/2} (\boldsymbol k) & =\begin{bmatrix}
-\frac{\sqrt{3}\, e^{-3 \,i\, \phi }  
\left(1-\cos^2 \theta \right) 
 \csc \left(\frac{\theta }{2}\right)} {4}  &
\frac{e^{-2 \,i \,\phi } \cos \left(\frac{\theta }{2}\right) (3 \cos  \theta  -1)}{2}&
   \frac{ e^{-i \,\phi }  \sin \left(\frac{\theta
   }{2}\right) \,(3 \cos  \theta  + 1)}{2}   &
   \frac{\sqrt{3} \, \sin \left(\frac{\theta }{2}\right) \sin  \theta  }  {2}  
\end{bmatrix}^{\rm T} 
\nn   \text{and }
\psi_{3/2} (\boldsymbol k) & = \begin{bmatrix}
e^{-3 \,i \,\phi } \cos^3\left(\frac{\theta }{2}\right) &
\frac{ \sqrt{3} \,e^{-2\, i \,\phi }
\sin^2 \theta  \csc \left(\frac{\theta }{2}\right)}{4}   &
   \frac{ \sqrt{3} \,e^{-i\, \phi } \sin \left(\frac{\theta }{2}\right) \sin  \theta}{2}   &
\sin^3\left(\frac{\theta }{2}\right)
\end{bmatrix}^{\rm T} .
\end{align} 
This will be used to determine the strength of the scattering cross-sections and, also, an ansatz for the nonequilibrium quasiparticle-distributions.

\subsection{Topological quantities}

Due to a nontrivial topology of the bandstructure, we need to compute the BC and the OMM, using the starting expressions of \cite{xiao_review,xiao07_valley}
\begin{align} 
\label{eqomm}
& {\boldsymbol \Omega}_s ( \boldsymbol k)  = 
    i  \left[ \nabla_{ \boldsymbol k}  \psi_{s}({ \boldsymbol k}) \right ]^\dagger 
    \cross  \left [ \nabla_{ \boldsymbol k}  \psi_{s}({ \boldsymbol k}) \right ]
\text{and } 
{\boldsymbol {m}}^s ( \boldsymbol k) = 
\frac{  -\,i \, e} {2 } \,
 \left[ \nabla_{ \boldsymbol k} \psi_{s} ({ \boldsymbol k}) \right ]^\dagger 
 \cross
\Big [
\left \lbrace \mathcal{H} ({ \boldsymbol k}) -\varepsilon^s
({ \boldsymbol k}) 
\right \rbrace
\left \lbrace \boldsymbol \nabla_{ \boldsymbol k} \psi_{s}({ \boldsymbol k})
 \right \rbrace \Big  ] ,
\end{align}
respectively. Evaluating these expressions for the RSW node described by $\mathcal{H}(\boldsymbol{k}) $, we get \cite{graf_thesis, ips-rsw-ph}
\begin{align}  
{\boldsymbol \Omega}_s ( \boldsymbol k) &=    
	 \frac{  - \,s \,\boldsymbol k  } {k^3}  \text{ and } 
\boldsymbol{m}_s( \boldsymbol k) 
= \frac{  - \,e \,  v_F \, \mathcal{G}_s  \,\boldsymbol k } {k^2} \, ,
\text{ where } \lbrace \mathcal{G}_{\pm 1/2}, \,\mathcal{G}_{\pm 3/2} \rbrace
 = \left \lbrace \frac{7}{4}, \, \frac{3}{4} \right \rbrace.
\end{align}
Since $\boldsymbol  \Omega_s ( \boldsymbol k)$ and $\boldsymbol{m}_s( \boldsymbol k)$ are the intrinsic properties of the bandstructure, they depend only on the wavefunctions. 

A nonzero BC manifests itself primarily via modifying the phase-space volume element, incorporated through the factor of \cite{mermin, sundaram99_wavepacket, li2023_planar, ips-kush-review, ips-rsw-ph, ips-shreya}
\begin{align}
{\mathcal D}_{s}  (\boldsymbol k) = \left [1 
+ e \,  \left \lbrace 
{\boldsymbol B} \cdot \boldsymbol{\Omega}_s  (\boldsymbol k)
\right \rbrace  \right ]^{-1}.
\end{align}
It also affects the explicit forms of the Hamilton's equations, which we will discuss in Sec.~\ref{secboltz}.
On the other hand, the OMM distorts the Fermi surfaces along the direction of $\boldsymbol B$, by adding the term
\begin{align}
\label{eqmodi}
\varepsilon_s^{(m)}   (\boldsymbol k) 
= -\, {\boldsymbol B} \cdot \boldsymbol{m}_s  (\boldsymbol k) \,,
\end{align}
as schematically depicted in Fig.~\ref{figdis}(b). Consequently, in addition to affecting the Fermi-Dirac-distribution functions, it causes modifications of the group-velocities via adding
\begin{align}
 {\boldsymbol   v}_s^{(m)} ({\boldsymbol k} ) \equiv 
 \nabla_{\boldsymbol k} \varepsilon_{(\sigma)}   (\boldsymbol k) 
\end{align}
to $\boldsymbol{v}_{s}(\boldsymbol{k})$.

\section{Conductivity}
\label{secboltz}

In this section, we will review the step-by-step procedure to compute the electric conductivity, applying the machinery of semiclassical Boltzmann equations \cite{mermin, sundaram99_wavepacket, li2023_planar, ips-kush-review, ips-rsw-ph, ips-shreya, timm, ips-exact-spin1}. Although this has been demonstrated multiple times in our earlier works \cite{ips-exact-spin1, ips-exact-kwn}, it is primarily needed here to set the notations. The main line of arguments for obtaining the final equations was developed in Ref.~\cite{timm}, which we adapt and generalise for our system under study. For the sake of simplicity, we will perform our calculations by setting $ T = 0$, and an application of a positive chemical potential, $\mu>0$, to the node. Consequently, the bands with $s=1/2$ and $s=3/2$ will contribute to transport.

\subsection{Kinetic equations driven by electromagnetic fields}

In thermal equilibrium, the distribution of the quasiparticles follows the Fermi-Dirac distribution,
\begin{align}
\label{eqdist}
f_0 \big (\xi_s (\boldsymbol k) , \mu, T \big )
= \left[ 1 + \exp \left \lbrace  
\left(  \xi_s (\boldsymbol k)-\mu \right) /T  \right \rbrace \right ]^{-1}, 
\quad \xi_s ({ \boldsymbol k}) = {\varepsilon}_s  ({ \boldsymbol k})  
+ \varepsilon_s^{(m)}   (\boldsymbol k) \,,
\end{align}
where $T $ is the temperature and
$ \xi_s ({ \boldsymbol k}) $ is the effective energy (containing the OMM contribution).
The quantity $f_0$ will appear in the equations that follow, where we will suppress showing its $\mu$- and $ T $-dependence explicitly, just for the sake of uncluttering of notations. We note that, as $T \rightarrow 0$, $\partial_u f_0 (u) \rightarrow - \delta (u -\mu ) $, which serves to reduce the momentum-space integrals over a Fermi pocket to the respective Fermi surface.

The Hamilton's equations of motion for the quasiparticles, occupying a given Bloch band, are described by \cite{mermin, sundaram99_wavepacket, li2023_planar, ips-kush-review, ips-rsw-ph, ips-shreya}
\begin{align}
\label{eqrkdot}
\dot {\boldsymbol r} & = \nabla_{\boldsymbol k} \, \xi_s  
- \dot{\boldsymbol k} \, \cross \, \boldsymbol \Omega_s  (\boldsymbol k)
\Rightarrow  \, \dot{\boldsymbol r}  = \mathcal{D}_s ({\boldsymbol k})
	\left[   \boldsymbol{w}_s ({\boldsymbol k}) +
	 e \, {\boldsymbol E}  \cross  
\boldsymbol \Omega_s ({\boldsymbol k})   + 
	 e   \left \lbrace \boldsymbol \Omega_s (\boldsymbol k ) \cdot \boldsymbol{w}_s ({\boldsymbol k}) 
 \right \rbrace  \boldsymbol B  \right], \nn & 
\text{and } 
\dot{\boldsymbol k} = -\, e  \left( {\boldsymbol E}  
+ \dot{\boldsymbol r} \, \cross\, {\boldsymbol B} 
	\right ) \Rightarrow
\dot{\boldsymbol k}  = -\, e \,\mathcal{D}_s ({\boldsymbol k})
  \left[   {\boldsymbol E} 
+   \boldsymbol{w}_s ({\boldsymbol k}) \cross  {\boldsymbol B} 
+  e  \left (  {\boldsymbol E}\cdot  {\boldsymbol B} \right )  
\boldsymbol \Omega_s  ({\boldsymbol k})\right],
\end{align}
where
\begin{align}
\boldsymbol{w}_s ({\boldsymbol k}) = {\boldsymbol  v}_s(\boldsymbol k )
+  {\boldsymbol   v}_s^{(m)} ({\boldsymbol k} ) \,.
\end{align}
The presence of various terms involving the BC and the OMM reflect the nontrivial role played by the topological properties, as compared to systems where such topology does not exist. For comparison, one can look at the structure of the phase-space equations in Refs.~\cite{ips-kush, ips_tilted_dirac}, for example. We notice that the term appearing as $- \dot{\boldsymbol k} \, \cross \, \boldsymbol \Omega_s $ represents an anomalous velocity, with the BC (existing in the momentum space) playing the counterpart of the magnetic field (existing in the real space). 

Let us provide an example regarding how to estimate the bounds on the allowed strength of $\boldsymbol{B}$ \cite{timm}.
For this purpose, we consider an RSWN subjected to a magnetic field oriented along the $z$-direction. Solving the the energy dispersion relation characterising the $n_L$-th Landau level (LL), we pick a level $ n_L $ carrying positive energy (i.e., $  n_L > 0$), which takes the form of
$ \epsilon^s_{n_L}(k_z) = |s|\, v_F \,\sqrt{2\,e\, B\, { n_L} + (v_F \,k_z)^2}$. For the semiclassical treatment to remain self-consistent, the number of occupied LLs, $N_L$, must be sufficiently large (i.e.,  $N_L \gg 1$). Put differently, the energy gap, $
\Delta \epsilon_s (B) \equiv |s|\, v_F \left( \sqrt{2 \,e\, B\,( {N}_L +1)} -\sqrt{2\,e\,B\, N_L} \right) $, separating the topmost filled LL from the lowest empty one at $k_z = 0$, must remain negligibly small in comparison with the chemical potential: $\Delta\epsilon_s (B) \ll \mu.$
A straightforward estimate of this energy gap proceeds as follows:
\begin{align}
 \Delta \epsilon_s (B) = |s|\,v_F \,\sqrt{2\,e \, B\, N_L }\;
 \left(\sqrt{1 + \frac{1}{N_L}} - 1\right) 
 \simeq |s|\, v_F\sqrt{\frac{e\,B}{2\, N_L}}.
\label{eqS4}
\end{align}
Noting that $N_L$ denotes the index of the highest-occupied LL, the chemical potential ($\mu$) must satisfy $ |s|\,v_F \, \sqrt{2 \, e \,B\,N_L} \leq \mu < |s|\, v_F \sqrt{2\,e\,B\,(N_L+1)}$ and
$  N_L = \left\lfloor \frac{1}{2\,e\,B}\left(\frac{\mu} {|s|\,v_F}\right)^2 \right\rfloor $,
where $\lfloor u \rfloor$ denotes the floor function (i.e., the largest integer not exceeding $u$). Combining everything then yields the criterion,
\begin{align}
    B \ll \frac{1}{e}\left(\frac{\mu} {|s|\, v_F}\right)^2,
    \label{S7}
\end{align}
which must be satisfied for the applied magnetic field to be regarded as weak and for the validity of the semiclassical approximation to be ensured. We have used $\mu = 0.1$ eV and $v_F = 6\times 10^{-6}$, which lead to an upper limit of the order of $10000$ eV$^2$, which is well above the ranges of $B$ used in our plots.

The semiclassical Boltzmann's transport formalism is implemented by starting with the basic kinetic equations, embodied by
\begin{align}
\label{eqkin32}
& \left [ \partial_t  
+ {\boldsymbol w}_s ({\boldsymbol k}) \cdot \nabla_{\boldsymbol r} 
-e \left \lbrace  \boldsymbol{E}
+  {\boldsymbol w}_s ({\boldsymbol k})  \times {\boldsymbol B} 
\right \rbrace  \cdot 
\nabla_{\boldsymbol k} \right ] f_s (\boldsymbol r, \boldsymbol k, t)
=  I_{\text{coll}} [f_s ({\boldsymbol k}) ]\,.
\end{align}
The function $f_s  (\boldsymbol r, \boldsymbol k, t) $ represents the nonequilibrium quasiparticle-distribution for the fermions populating band $s$, caused by a small deviation from $ f_0 (\xi_s(\boldsymbol k) ) $, as soon as probe fields (like electric field, temperature gradient \cite{ips-ruiz, ips-rsw-ph, ips-spin1-ph}, or chemical potential gradient \cite{ips-shreya}) are applied externally. On the right-hand side of Eq.~\eqref{eqkin32}, we have the \textit{collision integral}, $ I_{\text{coll}} [f_s ({\boldsymbol k}) ] $. Obviously, its role is to account for the appropriate scattering processes relaxing $f_s ({\boldsymbol k})$ towards the equilibrium value of $f_0 (\xi_s({\boldsymbol k}))$. 
As seen from the equation itself, the probe field here is just an electric field ($\boldsymbol E $), and the
deviation, $ \delta f_s  (\boldsymbol r, \boldsymbol k, t) \equiv 
f_s  (\boldsymbol r, \boldsymbol k, t) - f_0 (\xi_s (\boldsymbol k) ) $, is assumed to be of the same order of smallness (say, $\delta $) as $|\boldsymbol E|$. Since we restrict ourselves to spatially-uniform and time-independent electromagnetic fields, $f_s $ must must also be independent of position and time, so that $ \delta f_s  (\boldsymbol r, \boldsymbol k, t) = \delta f_s  (\boldsymbol k) $. Observing that $ \delta f_s  (\boldsymbol k) \propto |\boldsymbol E| \propto \delta $, we expand the terms on both sides upto a given order in $\delta $. When we are interested in the \textit{linear response}, we just retain the leading-order corrections, meaning terms which are linear-in-$\delta $.

In this paper, we will be focussing on the simple case of elastic point-scattering mechanisms, for which the collision integral takes the form of
\begin{align}
I_{\text{coll}} [f_s({\boldsymbol k})]
 = \sum \limits_{\tilde s}
\int_{k'}
\mathcal{M}_{ s, \tilde s} (\boldsymbol k,\boldsymbol k^\prime)
\left[ f_{\tilde s} ( \boldsymbol k^\prime))  - f_s (\boldsymbol k) \right ] ,
\text{ where  }
\int_{k^\prime} \equiv \int  \frac{ d^3 \boldsymbol k^\prime \, \mathcal{D}^{-1}_{\tilde s} ({\boldsymbol k}^\prime) } 
{ (2\,\pi)^3 }
\end{align}
is the compact symbol for indicating a 3d momentum-space integral. Needless to say, it includes the modified phase-space factor caused by a nonzero BC. The scatterings being elastic, they do not involve any dissipation of energy. Using the Fermi's golden rule, they can be expressed as
\begin{align}
\label{eqMimp}
\mathcal{M}_{ s, \tilde s} (\boldsymbol k,\boldsymbol k^\prime) 
= \frac{2\, \pi \, \rho_{\rm imp} } {V} \,
\Big \vert \left \lbrace  \psi_{\tilde s }({ \boldsymbol k^\prime }) 
\right \rbrace^\dagger 
\; {\mathcal V} _{ s, \tilde s} (\boldsymbol k,\boldsymbol k^\prime) 
  \;  \psi_s ({ \boldsymbol k}) \Big \vert^2 \,
 \delta \Big( \xi_{\tilde s } (\boldsymbol k^\prime)- \xi_s (\boldsymbol k ) \Big) \,,
\end{align}
where, $ \rho_{\rm imp}$ denotes the impurity-concentration, $V$ stands for the system's spatial volume, and ${\mathcal V} _{ s, \tilde s} (\boldsymbol k,\boldsymbol k^\prime) $ represents the matrix-elements of the effective scattering-potential. The consideration of elastic and (pseudo)spinless scattering-centres reduces the potential to an identity matrix in the spinor space, i.e., $ {\mathcal V} _{ s, \tilde s} (\boldsymbol k,\boldsymbol k^\prime) 
=  \mathbb{I}_{4 \times 4} \, {\mathcal V} _{ s, \tilde s} $. Thus, we can now use the version of
\begin{align}
\label{eqMimp1}
\mathcal{M}_{ s, \tilde s} (\boldsymbol k,\boldsymbol k^\prime) 
= \frac{2\, \pi \, \rho_{\rm imp} 
\, |{\mathcal V} _{ s, \tilde s} |^2} 
{V} \,
\Big \vert \left \lbrace  \psi_{\tilde s }({ \boldsymbol k^\prime }) \right \rbrace^\dagger 
\; \psi_{s}({ \boldsymbol k}) \Big \vert^2 \,
 \delta \Big( \xi_{\tilde \chi , \tilde s } (\boldsymbol k^\prime)
 - \xi_{  \chi , s } (\boldsymbol k ) \Big) ,
\end{align}
which can be easily computed by using the spinorial forms shown in Eq.~\eqref{eqev}. Assuming a reciprocal interband-scattering function between the $ s = 1/2$ and $s=3/2$ bands, we set $ {\mathcal V}_{ 1/2, 3/2 } = {\mathcal V}_{ 3/2, 1/2 }$. Under these assumptions, altogether we need three parameters for the scattering strengths, leading to
\begin{align}
\big | {\mathcal V}_{1/2,1/2} \big |^2  
   \equiv  \frac{ 16\times 2\,\pi} {\rho_{\rm imp}} \, \beta_{\rm intra}^{1/2,1/2} \,,\quad
\big | {\mathcal V}_{3/2,3/2} \big |^2  
   \equiv  \frac{ 16\times 2\,\pi} {\rho_{\rm imp}} \, \beta_{\rm intra}^{3/2,3/2} \,,   
 \text{ and } 
 \big | {\mathcal V}_{ s,{\tilde s} } \big |^2 \Big \vert_{ s \neq \tilde s} 
 \equiv \frac{ 16\times 2\,\pi} {\rho_{\rm imp}}\, \beta_{\rm inter} \,.
\end{align} 
We have added the self-explanatory subscripts, ``intra'' and ``inter'', to clearly indicate the intraband- and interband-parts, respectively.

\subsection{Linearised Boltzmann equation and its solutions}

Gathering all the ingredients of the preceding subsection, we are now ready to solve the \textit{linearised Boltzmann equation}, embodied by
\begin{align}
\label{eqkin5}
& - e\, {\mathcal D}_s ({\boldsymbol k})
 \left [
\left \lbrace {\boldsymbol{w}}_s ({\boldsymbol k})
+ e \,\Big(
{\boldsymbol \Omega}_s ({\boldsymbol k}) 
\cdot {\boldsymbol{w}}_s   ({\boldsymbol k}) \Big )  \boldsymbol B \right \rbrace
\cdot {\boldsymbol E}  
\; \;	\frac{\partial  f_0 (\xi_s({\boldsymbol k})) }
 {\partial \xi_s({\boldsymbol k}) }
+  
\left \lbrace   {\boldsymbol{w}}_s ({\boldsymbol k})
\cross  {\boldsymbol B}  \right \rbrace
\cdot \nabla_{\boldsymbol k}
\, \delta f_s (\boldsymbol k) \right] 
 =  I_{\text{coll}} [f_s ({\boldsymbol k})] \,.
\end{align}
Specifically, we choose the collinear electromagnetic fields to act along the $z$-axis, such that $\boldsymbol B  =  B\, \boldsymbol{\hat{z}}$ and $ \boldsymbol E  =  E\, \boldsymbol{\hat{z}}$.
To proceed further, we rewrite the deviation in the particles' distribution-functions as
\begin{align}
\label{eqansatz}
	\delta f_s (\boldsymbol {k}) =
-\,	 e\, 	\frac{\partial  f_0 (\xi_s) } {\partial \xi_s } 
	\,   {\boldsymbol E}  \cdot \bm{\Lambda}_s (\boldsymbol {k} )
=
-\,	 e\, 	\frac{\partial  f_0 (\xi_s({\boldsymbol k})) }
 {\partial \xi_s ({\boldsymbol k})} 
	\,   E \, {\Lambda}^z_s ( \boldsymbol {k} ) \,,
\end{align}
where $ \bm{\Lambda}_s (\boldsymbol {k} ) $ is the vectorial mean-free path. Because of our specific choice of coordinates, only the $z$-component of $ \bm{\Lambda}_s ( \boldsymbol {k} ) $ [i.e., $ {\Lambda}^z_s (\boldsymbol {k} )  $] emerges as the nontrivial component as a result of the probe electric field. Consequently, the only equation that needs to be solved turns out to be
\begin{align}
\label{eqvec}
&  w^z_s ({\boldsymbol k})
+ e \, B \left [
{\boldsymbol \Omega}_{s} ({\boldsymbol k})
 \cdot {\boldsymbol{w}}_s ({\boldsymbol k})  \right ]
-\, 
e \, B \left [ {\boldsymbol{w}}_s ({\boldsymbol k}) \cross   
\boldsymbol{\hat z}  ({\boldsymbol k}) \right ] 
\cdot \nabla_{\boldsymbol k} {\Lambda}^z_s (\boldsymbol {k} )  
 = {\mathcal D}^{-1}_s  ({\boldsymbol k})
\sum \limits_{\tilde \chi, \tilde s}
\int_{k'}
\mathcal{M}_{ s, \tilde s} (\boldsymbol k,\boldsymbol k^\prime)
\left[ 
 {\Lambda}^z_{\tilde s} (\boldsymbol {k}^\prime ) 
 -  {\Lambda}^z_s (\boldsymbol {k} )  \right ].
\end{align}
Looking at the structure of the terms on both the sides, we infer that
$ {\Lambda}^z_s \equiv {\Lambda}^z_s ( \mu, \theta)$ at an energy $\mu$, which is independent of $\phi $. The fact that $\delta f_s (\boldsymbol {k})$ can depend only on $\theta$ and $\mu$ follows from the constraints imposed by the delta functions, reducing an integral over the full momentum space to one over the Fermi-surface contour with energy-profile $\xi_s (\boldsymbol k) = \mu $. Essentially, the dependence on the azimuthal angle ($\phi$) drops out because of the residual rotational symmetry of the system about the $z$-axis. The equations of the Fermi surfaces are derived by solving $\xi_s 
(k_F^s , \theta )= \mu $, with the $\theta$-dependent radii of $ \lbrace k_F^s (\theta) \rbrace $ parametrising the local Fermi momenta. Overall, the justification of assuming the $\phi$-independent forms of $\lbrace {\Lambda}^z_s \equiv {\Lambda}^z_s ( \mu, \theta) \rbrace $ is evident from the self-consistency ---  $\left [ {\boldsymbol \Omega}_s (\boldsymbol k) \cdot
 {\boldsymbol{w}}_s (\boldsymbol k)  \right ] $ is $\phi$-independent and $ \left [  {\boldsymbol{w}}_s 
 (\boldsymbol k)  \cross   \boldsymbol{\hat z}  \right ] 
\cdot \nabla_{\boldsymbol k} {\Lambda}^z_s (\mu, \theta ) $ reduces to zero.

Making use of the above arguments, we can now compute the spinor overlaps, $\big \vert \left \lbrace  \psi_{\tilde s }({ \boldsymbol k^\prime }) \right \rbrace^\dagger \; \psi_{s}({ \boldsymbol k}) \big \vert^2 $, with the help of Eq.~\eqref{eqev}. Knowing that the dependence of the azimuthal angles, $\phi$ and $\phi^\prime $, will disappear once we perform the azimuthal-angle integrations, we list the
overlap-functions as
\begin{align}
\label{eqoverlap}
{\mathcal T }_{ s, \tilde s} (\theta, \theta^\prime)
& = \left[  5 -3 \cos^2 \theta^\prime 
+ \cos  \theta  \left(17 \cos  \theta^\prime  -
27 \cos^3 \theta^\prime  \right)+\cos^2 \theta  
\left(9 \cos^2 \theta^\prime  -3\right) 
  +9 \cos^3 \theta  \cos  \theta^\prime   \left(5 \cos^2 \theta^\prime  -3\right)
    \right ] \beta_{\rm intra}^{1,1} \nn & \quad
+ \left[ 5 - 3\cos^2 \theta^\prime
+\cos  \theta  \left(9 \cos\theta^\prime  -3 \cos^3 \theta^\prime  \right)
+ \cos^2 \theta  \left(9 \cos^2 \theta^\prime  -3\right)
+ \cos^3 \theta  \left(5 \cos^3 \theta^\prime  -3 \cos  \theta^\prime  \right)  
\right ] \beta_{\rm intra}^{3,3} \nn & \quad
+ \left[ 3 + 3\cos^2 \theta^\prime  
+ \cos^3 \theta  \left(9 \cos  \theta^\prime  -15 \cos^3 \theta^\prime  \right)
+\cos  \theta  \left(9 \cos^3 \theta^\prime  -3 \cos  \theta^\prime  \right)
   +\cos^2 \theta  \left(3-9 \cos^2 \theta^\prime  \right)  
 \right ]  \beta_{\rm inter}\,.
\end{align}
Here, the answer contains the results obtained after already performing the trivial $\phi$ and $ \phi'$ integrals. We are now ready to solve Eq.~\eqref{eqvec}, starting from its simplified version of
\begin{align}
\label{eq_lambda_mu}
h_s (\mu, \theta) = 
 \sum_{\tilde s} V  
\int_{k^\prime }\, \mathcal{M}_{ s, \tilde s} 
(\boldsymbol k,\boldsymbol k^\prime)
\, {\Lambda}^z_{\tilde s} ( \mu, \theta^\prime ) 
 - \frac{ {\Lambda}^z_s  (\mu, \theta)} 
{\tau_s(\mu, \theta)}  \,,
\end{align}
where
\begin{align}
\label{eqtau}
\tau^{-1}_s(\mu, \theta)
= \sum_{\tilde s} V \int_{k^\prime }
\mathcal{M}_{ s, \tilde s} (\boldsymbol k,\boldsymbol k^\prime) \,, \quad
h_s (\mu, \theta) = {\mathcal D}_s ({\boldsymbol k})
 \left[ w^z_s ({\boldsymbol k})
+ e \, B \left \lbrace
{\boldsymbol \Omega}_s ({\boldsymbol k}) \cdot {\boldsymbol{w}}_s ({\boldsymbol k})
  \right \rbrace \right].	
\end{align}
Incorporating the energy-conservation restrictions, which reduce the integrals to the respective Fermi surfaces at energy $ \mu $, we simplify it further to
\begin{align}
\label{eqlambdamu}
& h_s(\mu, \theta) + \frac{ {\Lambda}^z_s  (\mu, \theta)} 
{\tau_s(\mu, \theta)} =
\sum_{\tilde s}  
\frac{ \rho_{\rm imp} 
\, |{\mathcal V} _{ s, \tilde s} |^2 }
{ 16 \times 2\, \pi }
\int d\theta^\prime \, \frac{\sin \theta^\prime \left (k^\prime \right )^3 
\, {\mathcal D}^{-1}_{\tilde s} ({\boldsymbol k}^\prime)}
{  |\boldsymbol k^\prime \cdot {\boldsymbol{w}}_s (\boldsymbol k^\prime)  | }
\, {\mathcal T }_{ s, \tilde s} (\theta, \theta^\prime)
\, {\Lambda}^z_{\tilde s} (\mu, \theta^\prime ) 
\Big \vert_{ k^\prime = k_F^{\tilde s} } \,.
\end{align}
We note that the factor $ \sin \theta^\prime \left (k^\prime \right )^3  $ arises as the Jacobian for switching to the spherical-polar coordinates. Tackling the factors of $ \delta \Big( \xi_{\tilde s } (\boldsymbol k^\prime) -\mu \Big) $, arising from Eq.~\eqref{eqMimp1}, needs re-expressing them in the language of $\delta (k^\prime - k_F^{\tilde s})$, which sources the appearance of the other multiplicative factor of 
$$  \big |\boldsymbol {\hat k^\prime} \cdot 
 \nabla_{\boldsymbol k^\prime} \xi_s({\boldsymbol k^\prime}) \big |^{-1}  
 = k^\prime / |\boldsymbol k^\prime \cdot {\boldsymbol{w}}_s (\boldsymbol k^\prime)  |\,.$$

As a final step, observing the powers of $\cos \theta $ in \eqref{eqoverlap}], we set forth an ansatz of
\begin{align}
 {\Lambda}^z_s (\mu,\theta ) & = 
\tau_s (\mu,\theta)
 \left [ \lambda_s - h_s + a_s \cos \theta  + b_s \cos ^2 \theta 
 + c_s \cos^3 \theta   \right ].
\end{align}
In total we have 8 undetermined coefficients, denoted as the set $\lbrace \lambda_s , \, a_s , \, b_s , 
\, c_s \rbrace $. We have 8 linear equations at our disposal, extracted from Eq.~\eqref{eqlambdamu} by equating the
coefficients of $1$, $\cos \theta $, $\cos^2 \theta $, and $\cos^3 \theta $ on both the sides for each value of $s$.
This system of linear equations can be expressed compactly as a matrix equation, captured by
\begin{align}
\label{eqmatrix}
\mathcal A \, \mathcal N = \Upsilon \,, \text{ where }
\mathcal N =\begin{bmatrix}
\lambda_{1/2} & a_{1/2} & b_{1/2} & c_{1/2} & \lambda_{3/2} & a_{3/2} & b_{3/2} & c_{3/2}
\end{bmatrix}^{\rm T}\,.
\end{align}
The explicit forms of the square and row matrices, $ \mathcal A $ and $ \Upsilon $, are demonstrated in the appendix.
Now there is a catch --- the eight equations represented above are not linearly independent. This is because $ \mathcal A $'s rank is lower than 8. It is actually well and good since  it prevents from the system from becoming overdetermined, as we must also employ the constraint of the electron-number conservation, embodied by
\begin{align}
\label{eqcon}
\sum \limits_{s}	\int_k  \delta f_s (\boldsymbol {k}) = 0\,.
\end{align}

\begin{figure*}[]
\subfigure[]{\includegraphics[width= 0.75 \textwidth]{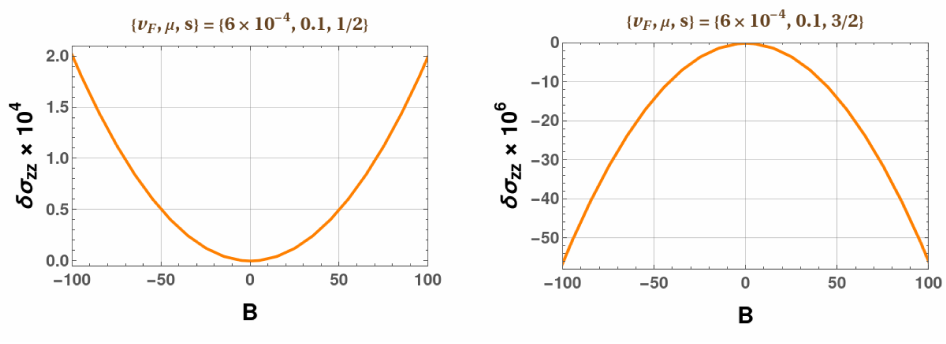}}
\subfigure[]{\includegraphics[width= 0.75 \textwidth]{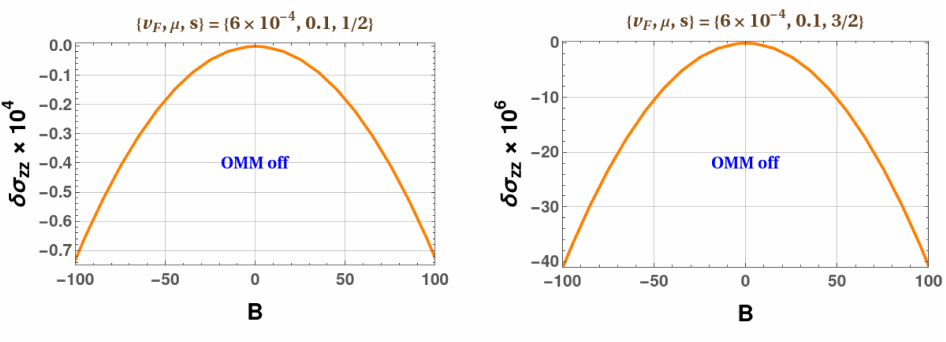}}
\caption{\label{figsep}$\delta \sigma_{zz} (s) $ from each of the bands considering no interband interactions, with $\beta_{\rm intra}^{s,s} $ set to unity: While subfigure (a) depicts the variation of the full conductivity with $B$ (in eV$^2$) when OMM is taken into account appropriately, subfigure (b) represents the conductivity versus $B$ characteristics when OMM is turned off. Here, $\mu $ is in eV.}
\end{figure*}

Plugging in the solutions obtained, the charge-current density along the $z$-direction is computed as
\begin{align}
\label{def_cur}
	J_z^{\rm tot} =- \,\frac{e}{ V }\sum_s
\int_k	\left[ {\mathcal D}_s ({\boldsymbol k}) \right ]^{-1}
(\dot{\boldsymbol {r}} \cdot \boldsymbol{\hat z})
\;  \delta f_s (\boldsymbol k)\,,
\end{align}
finally leading to the desired longitudinal magnetoconductivity,
\begin{align}
\sigma_{zz}^{\rm tot} = \sum_s \sigma_{zz}^s \,, \quad
 \sigma_{zz}^s
=  -\,\frac{e^2 } { V } 
\int \frac{d^3 {\boldsymbol k}} {(2\, \pi)^3} 
	\left[   w^z_s (\boldsymbol k)
+ e \, B  \left \lbrace \boldsymbol \Omega_s (\boldsymbol k)
\cdot 	  \boldsymbol{w}_s (\boldsymbol k) \right  \rbrace   \right] 
\delta \big (\xi_s (\boldsymbol k)-\mu \big) \, {\Lambda}^z_s (\mu, \theta )	\,. 
\end{align}

\begin{figure*}[]
\subfigure[]{\includegraphics[width = 0.75 \textwidth]{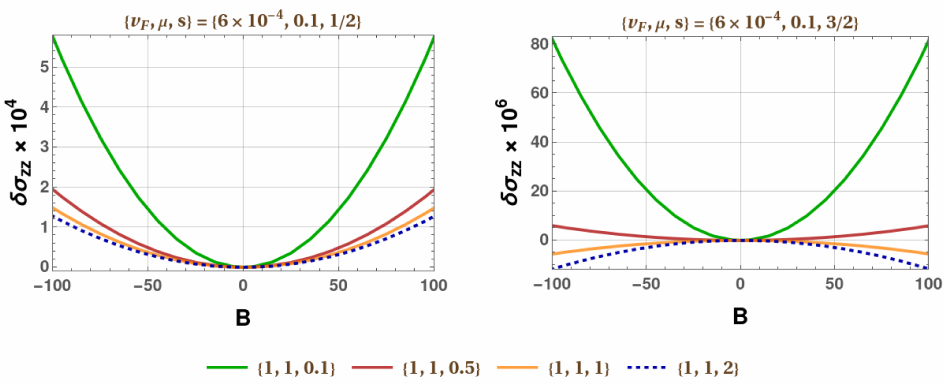}}
\subfigure[]{\includegraphics[width = 0.75 \textwidth]{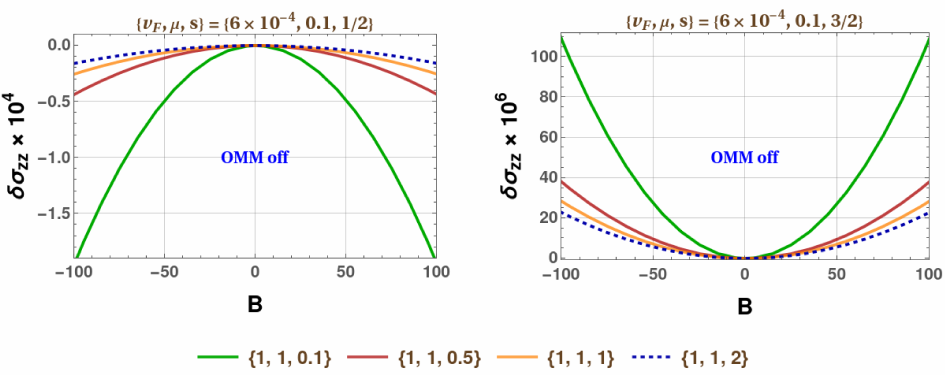}}
\caption{\label{figall}$\delta \sigma_{zz} (s) $ from each of the bands considering both intraband and interband interactions. While subfigure (a) depicts the variation of the full conductivity with respect to $B$ (in eV$^2$) when OMM is taken into account appropriately, subfigure (b) represents the conductivity-versus-$B$ profiles when OMM is switched off. The three numbers in each plot-legend indicate the values in the set $\lbrace \beta_{\rm intra}^{1/2,1/2},\; \beta_{\rm intra}^{3/2, 3/2},\; \beta_{\rm inter} \rbrace $. Here, $\mu $ is in eV.}
\end{figure*}

\subsection{Results and discussions} 
\label{secres}

The solutions of Eq.~\eqref{eqlambdamu} are obtained numerically for a given set of parameter-values. In our illustrations, we resort to plotting
$$\delta \sigma_{zz} (s) \equiv \sigma^s_{zz} (B) / \sigma^s_{zz}(B=0)  -1 ,$$
where the residual conductivity at $B=0$ (i.e.,  the Drude part) is subtracted off. The parameter values have been taken from Ref.~\cite{xu2020_optical}. There, $\hbar\, v_F = 1.23 $ eV {\AA}, which translates into
\begin{align*}
v_F  = \frac{1.23 \times 10^{-10} \text{ m eV}} {6.58 \times 10^{-16} \text{ eV s}}
= 0.1869  \times 10^{6}\text{ m/s} \simeq 0.06 \times 10^{-2} \,c\,,
\end{align*}
where $c$ is the speed of light. Therefore, in natural units, we set the value of $v_F$ to $6\times 10^{-4}$. In contemporary experiments, scientists have reported using magnetic fields upto $\sim 10 $ Tesla \cite{claudia-multifold}, which is equivalent to $\sim 1950$ eV$^2$ in natural units. Our representative plots depict the global behaviour of conductivity within this experimentally-feasible range.

To start with, we attempt to make sense of the response-characteristics for the scenarios when there is no interband scattering, i.e., $\beta_{\rm inter } $ is set to zero. This leads to the reduction of $  \mathcal A $ into the direct sum of the two matrices, $ \mathcal{A}_1 $ and $ \mathcal{A}_2 $, each of them being a $4\times 4$ matrix. In every case, we find that they have rank $3$ (instead of $4$), with $ \int_k  \delta f_s (\boldsymbol {k}) = 0$ providing the necessary extra equation. Fig.~\ref{figsep} captures such a scenario. Curiously, the parabolic-shaped response-profiles curve in opposite ways for $s=1/2$ and $s= 3/2$. In order to gauge the nature of the OMM-contributions, the lower panel represents the behaviour obtained by switching off the OMM. It leads to the conclusion that (I) the OMM-parts come with a positive sign and are strong enough to flip the BC-only response into the positive domain for $s=1/2$; (II) the OMM-contributions have a negative sign which assist in increasing the magnitude of the already negative-valued response (from the BC-only parts) for $s=3/2$.

Next, we consider the generic situations when the quasiparticles populating both the bands can scatter with each other. The $ 8 \times 8$  matrix $ \mathcal{A}$ is found to have rank $7 $, ensuring the internal consistency of the available equations. As discussed earlier, this means that the charge-conservation provides the remaining linearly independent equation [viz., Eq.~\eqref{eqcon}]. Fig.~\ref{figall} provides some representative curves. From the data presented here, as well as our scans of the parameter space (not shown here), we find that for $ \beta_{\rm inter} / \beta_{\rm intra}^{s,s} $ greater than some small threshold value, the curves for $s=3/2$ are turned upside down (into the negative domain), while the $s=1/2$ continuing to curve upwards (remaining positive). To probe into the role of the OMM-sourced parts, we plot the nature of the curves by switching off the OMM (lower panel of Fig.~\ref{figall}). Intriguingly, the curves for $s=1/2$ and $s=3/2$ behave in an opposite nature, the former always remaining negative and curving downward, irrespective of the values of $ \beta_{\rm inter} $. Overall, comparing the upper and the lower panels, we conclude that, while the OMM-parts have a positive sign and are strong enough to flip the BC-only response into the positive domain for $s=1/2$, the OMM-contributions have a negative sign and try to pull the positive-valued response (from the BC-only parts) towards lower values for $s=3/2$.

Overall, $\sigma^s_{zz} (B)$ comprises only even powers of $B$, which is in agreement with the Onsager-Casimir reciprocity relation, $\sigma^{\rm intra(inter)}_{zz} (\mathbf{B})= \sigma^{\rm intra(inter)}_{zz} (-\mathbf{B})$ \cite{onsager31_reciprocal, onsager2, onsager3}.
From analysing the generic features, we conclude that there is little wisdom to be gained from the $\beta_{\rm inter}  = 0 $ case to infer anything regarding the $\beta_{\rm inter}  \neq 0 $ cases. The only thing that matches between the two separate analyses is regarding the signs and strengths of the OMM-sourced parts. Albeit, the rationale behind not to try to extract similarities between the two cases is that the charge-conservation equations that come into picture are distinctly different --- (I) for a single band with $ \beta_{\rm inter}  = 0$, the charge-conservation must be satisfied individually for that particular band; (II) a nonzero interband-coupling demands that the net charge, from summing over the two bands, must be conserved. The observations serve to reinforce the recurrent phenomenon that the response from a single isolated band is fundamentally different from the case when the quasiparticles are free to scatter with those populating other bands. In fact, the same thing has been observed for the two bands of an isolated Kramers-Weyl node \cite{ips-exact-kwn}, where downward-curved parabolas morph into upward-curved ones for each of the bands, as soon as $\beta_{\rm inter}$ is turned on.

\section{Comparison with the results obtained from relaxation-time approximation}
\label{seccomp}

\begin{figure*}[]
\includegraphics[width = 0.75 \textwidth]{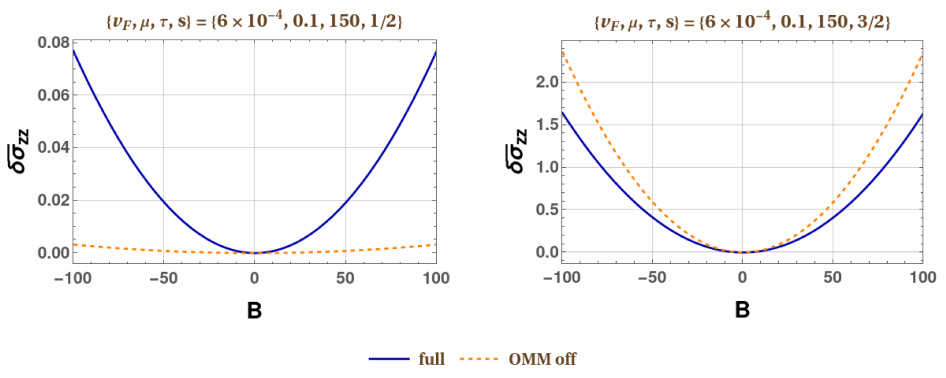}
\caption{\label{figrta}$\delta {\bar \sigma}_{zz} (s) $ from each of the bands, considering the RTA expression shown in Eq.~\eqref{eqrta}. While the dark-blue curves depict the variation of the full conductivity with respect to $B$ (in eV$^2$) when OMM is taken into account, the orange curves represent the conductivity-versus-$B$ profiles when OMM is switched off. Here, $\mu $ is in eV and $\tau$ is in eV$^{-1}$.}
\end{figure*}

In our earlier work \cite{ips-rsw-ph}, we obtained the results for planar-Hall conductivity via the naive application of the RTA. For the sake of comparison, we elaborate on the explicit expressions of the longitudinal magnetoconductivity reported therein. For a chemical potential $\mu>0$ and in the $T\rightarrow 0$ limit, the final result takes the following form:
\begin{align}
\label{eqrta}
& {\bar \sigma}^s_{zz}  =  \sigma^{s,\text{dr}}_{zz}
+  \sigma^{s,\text{bc}}_{zz} + \sigma^{s,m}_{zz}\,,\nn
& \sigma^{s,\text{dr}}_{zz}=  \frac{e^2 \, \tau\,\mu^2 } {6 \,\pi^2 \, s \, v_0}  \,,\quad
 \sigma^{s,\text{bc}}_{zz} =    \frac{ 4\, e^4 \, \tau \,  s^5 \,  v_0^3\,  B^2 } 
{ 15 \, \pi^2\,\mu^2 }   \,,
\quad  \sigma^{s,m}_{zz}  = 
 \frac{e^4 \, \tau \,  s \,  v_0^3 \, \mathcal{G}_s \,B^2}  {30 \, \pi^2 }  \, 
\frac{ 5 \, \mathcal{G}_s - 9 \, s^2  } {\mu^2} \,.
\end{align}
Here, $\tau$ is the momentum-independent relaxation time within the RTA framework and the superscripts, ``dr'', ``bc'', and ``$m$'', indicate the Drude (i.e., $B$-independent), BC-only, and OMM-induced contributions, respectively.
Adding up the three parts, we get
\begin{align} 
\label{eqrtatot}
{\bar \sigma}^s_{zz} 
=  \frac{e^2 \, \tau \,\mu^2}{6 \, \pi^2 \, s\, v_0} 
+   \frac{e^4 \, \tau \,  {s} \,  v_0^3\,  B^2} {30 \, \pi^2 }   \,
\frac{ 8 \, s^4 -9 \,s^2\, \mathcal{G}_s + 5 \,\mathcal{G}_s^2 } {\mu^2} \, .
\end{align}
We note that the two conduction bands (viz. for $s=1/2$ and $s=3/2$) at the node at cut by the same chemical potential and, so, there is no scope of having interband scatterings, unlike when we employ beyond-the-RTA (aka exact) method. So the net nodal contribution just comes from the sum of ${\bar \sigma}^{s=1/2}_{zz}$ and ${\bar \sigma}^{s=3/2}_{zz}$. We find that $\sigma^{s,\text{bc}}_{zz} $ is positive for both bands. On the other hand, $\sigma^{s,m}_{zz}$ is positive for $s=1/2$ and negative for $s=3/2$.

Let us define
\begin{align}
\delta {\bar \sigma}_{zz} (s) \equiv \bar \sigma^s_{zz} (B) / \sigma^{s,\rm dr}_{zz} - 1 \,.
\end{align}
For Fig.~\ref{figrta} shows the $\delta {\bar \sigma}_{zz}$-curves for the two bands, using the same parameter values as in Fig.~\ref{figsep} and setting $\tau =  150 $ eV$^{-1}$. Comparing the RTA curves with the exact-calculation curves for $\beta_{\rm inter} = 0$ (viz. Figs.~\ref{figrta} and \ref{figsep}), we infer that the naive RTA approach fails to capture the actual characteristics that have been obtained by the exact solutions
of the Boltzmann equations. We summarise our observations below:
\begin{enumerate}
\item For $s=1/2$, the total conductivity has the same sign but values differing by orders of magnitude. The more dramatic effect is seen when OMM is not included --- the curves are flipped. In the RTA case, the BC-only and OMM-induced parts have the same sign, with the magnitude of the latter dominating over the former. Whereas we have found that, in the exact-calculation case, the BC-only and OMM-induced parts have opposite signs, but with the magnitude of the latter dominating over the former.

\item For $s=3/2$, the total conductivity itself is seen to have opposite signs for the two methods. For the exact-solution case, the BC-only and the OMM-induced parts are both negative and reinforce each other. As discussed earlier, for the RTA results, the BC-only and the OMM-induced parts are positive and negative, respectively, with the magnitude of the former dominating over the latter. So, in this case, the RTA is seen to miserably fail even to produce any trustworthy qualitative features. 

\item We find that the interband scatterings, which are possible to incorporate only with the exact-solution method, change the single-band results drastically, introducing entirely new features starting at the qualitative level. Saliently, depending on the value of the ratio, $ \beta_{\rm inter} / \beta_{\rm intra}^{s,s} $, the parabolic conductivity-curves for $s=3/2$ can be oriented upward or downward [cf. Fig.~\ref{figall}(a)]. Of course, such drastic variations cannot be observed from the RTA calculations, due to the absence of the possibility of interband scatterings.

\end{enumerate}

\section{Summary and outlook}
\label{secsum}

In the current piece of work, we have ventured into determining the nature of the linear response in the form of longitudinal magnetoconductivity, focussing on an isolated RSWN. We have attempted to include both the BC and the OMM in a cohesive manner, as the necessary topological ingredients in the semiclassical Boltzmann equations. A major stride forward  has been to solve the linearised (in the strength of the probe field) equations exactly, not limiting to the RTA with a constant phenomenological time ($\tau$). Needless to say, our results quantify the serious drawbacks of using momentum-independent relaxation times and neglecting the angle-dependent spinor-overlaps in the collision integrals for the analytical expressions). By comparing the nontrivial results with those obtained in our earlier works~\cite{ips-rsw-ph, ips-internode}, we do identify the inadequacies of dropping the angular-dependence in $\tau$ and the ramifications of the nontrivial spinor-structures, as the exact answers turn out to be fundamentally different.
Of course, in Ref.~\cite{ips-rsw-ph}, we did not account for interband scatterings, and neither shall we attempt to do so anymore, because we have already successfully implemented the generic formalism without resorting to any approximation. Another point is that, a linear-in-momentum term accompanying the identity matrix (in the operator space for the four-component spinors), is not allowed in the most generic Hamiltonian for the RSWNs~\cite{xu2020_optical} derived from group-theoretical arguments. Consequently, tilting is completely irrelevant for the RSWNs, preventing the possibility of having odd powers of $B$. We note that tilting is a ubiquitous feature of the WSMs \cite{staalhammar20_magneto, yadav23_magneto, rahul-jpcm, ips-ruiz, ips-tilted}, because the conventional Weyl nodes generically appear at the low-symmetry points of the BZs (not limited to those of chiral crystals). 

We would like to emphasise that the response for an RSWN is distinct from a mere superposition of conventional Weyl contributions. In particular, the multifold character has introduced the possibility of interband scatterings (not captured within the RTA framework), which is manifested by the conductivity not being equal to the sum of the curves with $\beta_{\rm inter}$ set to zero. Crucially, depending on the dominant scattering mechanism, we can tune the value of $ \beta_{\rm inter} / \beta_{\rm intra}^{s,s} $, which can lead to differing shapes of the the conductivity curves, as observed in Fig.~\ref{figall}(a). Such a possibility does not arise for the case of WSMs, which were studied in Ref.~\cite{timm}. As for the question of distinctive response from different types of multifold nodes, the eigenspinors and the topological quantities will generically be different, which will give rise to different characteristics of the conductivity curves, both qualitatively and quantitatively. In particular, the difference in eigenspinors gives rise to distinct forms of the quantity $ \mathcal{M}_{ s, \tilde s} $ appearing in the collision integrals, leading to distinct functional forms of the overlap functions [viz. Eq.~\eqref{eqoverlap}] and the relaxation-times as well [viz. Eq.~\eqref{eqtau}]. Two such distinct examples can be readily found in Refs.~\cite{ips-exact-spin1, ips-exact-kwn}.

In crystal-lattices, it is essential to satisfy the Nielsen-Ninomiya theorem \cite{nielsen81_no}, which is physically reflected by the fact that the Weyl nodes (with opposite chiralities) must always exist in pairs in the entire BZ. For the spinless WSMs, arising in achiral crystals (e.g., TaAs family \cite{lv_Weyl}), such conjugate partners are typically (almost) degenerate in energy, due to the presence of mirror or other roto-inversion symmetries. Therein, charge-pumping is an important internode phenomenon, in the presence of external electric ($\boldsymbol E $) and magnetic ($\boldsymbol B $) fields, having nonvanishing collinear components. This is an embodiment of the chiral anomaly in the arena of condensed matter physics \cite{chiral_ABJ, son13_chiral, li_nmr17, ips-internode}, where the analog of the well-known concept of spacetime chirality (of relativistic fermions) is realised in the momentum space. On the contrary, the oppositely-charged chiral nodes in chiral crystals need not be degenerate in energy, because the conjugate nodes are not related by crystal symmetries. In fact, they are observed to have discernible separations in energy and momenta, with an isolated RSWN located at an intrinsic chemical potential. As such, internode-scattering-induced charge-pumping becomes unimportant, while enhancing other phenomena like quantised circular photogalvanic effect \cite{bernevig, chang2018, ni2021_giant, moore18_optical, guo23_light,kozii, ips_cpge} and circular dichroism \cite{ips-cd1, ips_cd}. Thus, a single node will be relevant here, when we tune the external chemical potential to lie near the intrinsic energy of the concerned nodal point, as the chirally-conjugate node is well-separated in energy and momentum (see, for example, Refs. \cite{titus-ksm, law-ksm}). In other words, the energy-separation between the $\chi =\pm $ members of the pair is much larger than the temperature-scale and the chemical-potential variation of the experiments. Hence, as long as we stick to a variation of $\mu$ within $\sim 0.2 $ eV, our results are robust, since internode scatterings will be inconsequential.

In the future, it will be a rewarding enterprise to redo the calculations of linear response at finite temperatures \cite{ips_rahul_ph_strain, rahul-jpcm, ips-ruiz, ips-shreya, ips-rsw-ph, ips_tilted_dirac, ips-spin1-ph, ips-hermann-review}, which will also enable us to compute the temperature-dependent linear-response coefficients like the thermoelectric and thermal-conductivity tensors. Another avenue worth exploring is to investigate the effects of strain \cite{awadhesh_rsw-strain, ips_rahul_ph_strain, ips-ruiz, ips-rsw-ph}, which breaks the rotational symmetry and can lead to pseudomagnetic fields ($\boldsymbol B_5 $). A nonzero $\boldsymbol B_5$ makes it possible for odd powers of $ B$ to show up in the response.

\section*{Acknowledgments}

This research, leading to the results reported, has received funding from the Council of Science \& Technology (CST), U.P.

\section*{Data availability}

All data generated or analysed during this study are available within the article.

\appendix

\section*{Appendix: Explicit forms of the matrices}
\label{appmat}

The two matrices, $\mathcal A$ and $ \Upsilon $, shown in Eq.~\eqref{eqmatrix}, take the explicit forms of
{\tiny
\begin{align*}
&  \left[ 
\begin{matrix}
1 +3 \, c_1^2 \, \beta_{\text{intra}}^{1/2, 1/2}-5 \, c_1^0 \, \beta_{\text{intra}}^{1/2, 1/2}
& 3 \, c_1^3 \, \beta_{\text{intra}}^{1/2, 1/2}-5 \, c_1^1 \, \beta_{\text{intra}}^{1/2, 1/2} 
   & 3 \, c_1^4 \, \beta_{\text{intra}}^{1/2, 1/2}-
   5 \, c_1^2 \, \beta_{\text{intra}}^{1/2, 1/2} 
   & 3 \, c_1^5 \, \beta_{\text{intra}}^{1/2, 1/2}-
   5 \, c_1^3 \, \beta_{\text{intra}}^{1/2, 1/2} \\ \\
 27 \, c_1^3 \, \beta_{\text{intra}}^{1/2, 1/2}-
 17 \, c_1^1 \,\beta_{\text{intra}}^{1/2, 1/2} 
 & 1 + 27 \, c_1^4 \, \beta_{\text{intra}}^{1/2, 1/2}
 -17 \, c_1^2 \, \beta_{\text{intra}}^{1/2, 1/2}
  & 27 \, c_1^5 \, \beta_{\text{intra}}^{1/2, 1/2}
   -17 \, c_1^3 \, \beta_{\text{intra}}^{1/2, 1/2} 
   & 27 \, c_1^6 \, \beta_{\text{intra}}^{1/2, 1/2}
   -17 \, c_1^4 \, \beta_{\text{intra}}^{1/2, 1/2} \\ \\
 3 \, c_1^0 \, \beta_{\text{intra}}^{1/2, 1/2}
 -9 \, c_1^2 \, \beta_{\text{intra}}^{1/2, 1/2} 
 & 3 \, c_1^1 \,\beta_{\text{intra}}^{1/2, 1/2}
 -9 \, c_1^3 \, \beta_{\text{intra}}^{1/2, 1/2} 
   & 1-9 \, c_1^4 \, \beta_{\text{intra}}^{1/2, 1/2}
   +3 \, c_1^2 \, \beta_{\text{intra}}^{1/2, 1/2}
   & 3 \, c_1^3 \, \beta_{\text{intra}}^{1/2, 1/2}
   -9 \, c_1^5 \, \beta_{\text{intra}}^{1/2, 1/2} \\ \\
 27 \, c_1^1 \,\beta_{\text{intra}}^{1/2, 1/2}
 -45 \, c_1^3 \, \beta_{\text{intra}}^{1/2, 1/2} 
 & 27 \, c_1^2 \, \beta_{\text{intra}}^{1/2, 1/2}
 -45  \,  c_1^4 \, \beta_{\text{intra}}^{1/2, 1/2} 
   & 27 \, c_1^3 \, \beta_{\text{intra}}^{1/2, 1/2}
 -45 \, c_1^5 \, \beta_{\text{intra}}^{1/2, 1/2} 
   & 1-45 \, c_1^6 \, \beta_{\text{intra}}^{1/2, 1/2}
   +27 \, c_1^4 \, \beta_{\text{intra}}^{1/2, 1/2} \\ \\
 -3 \, c_1^2 \, \beta_{\text{inter}}-3 \, c_1^0 \, \beta_{\text{inter}} 
 & -3 \, c_1^3 \, \beta_{\text{inter}}-3 \, c_1^1 \,\beta
   _{\text{inter}} & -3 \, c_1^4 \, \beta_{\text{inter}}
   -3 \, c_1^2 \, \beta_{\text{inter}} 
   & -3 \, c_1^5 \, \beta_{\text{inter}}
   -3 \, c_1^3 \, \beta_{\text{inter}} \\ \\
 3 \, c_1^1 \,\beta_{\text{inter}}-9 \, c_1^3 \, \beta_{\text{inter}} 
 & 3 \, c_1^2 \, \beta_{\text{inter}}-9 \, c_1^4 \, \beta_{\text{inter}} 
   & 3 \, c_1^3 \, \beta_{\text{inter}}-9 \, c_1^5 \, \beta_{\text{inter}} 
   & 3 \, c_1^4 \, \beta_{\text{inter}}-9 \, c_1^6 \, \beta_{\text{inter}} \\ \\
 9 \, c_1^2 \, \beta_{\text{inter}}-3 \, c_1^0 \, \beta_{\text{inter}} 
 & 9 \, c_1^3 \, \beta_{\text{inter}}-3 \, c_1^1 \,\beta_{\text{inter}} 
 & 9 \, c_1^4 \, \beta_{\text{inter}}
   -3 \, c_1^2 \, \beta_{\text{inter}} 
   & 9 \, c_1^5 \, \beta_{\text{inter}}-3 \, c_1^3 \, \beta_{\text{inter}} \\ \\
 15 \, c_1^3 \, \beta_{\text{inter}}-9 \, c_1^1 \,\beta_{\text{inter}} 
 & 15 \, c_1^4 \, \beta_{\text{inter}}-9 \, c_1^2 \, \beta
   _{\text{inter}} & 15 \, c_1^5 \, \beta_{\text{inter}}-9 \, c_1^3 \, \beta_{\text{inter}} 
 & 15 \, c_1^6 \, \beta_{\text{inter}}-9 \, c_1^4 \, \beta_{\text{inter}} \\\\
\end{matrix}\right.\nn
& \hspace{ 3.5 cm}
\left.\begin{matrix}   
 -3 \, c_2^2 \, \beta_{\text{inter}}-3 \, c_2^0 \, \beta_{\text{inter}}  
 & -3 \, c_2^3 \, \beta_{\text{inter}}-3 \, c_2^1 \, \beta_{\text{inter}} 
 & -3 \, c_2^4 \, \beta_{\text{inter}}-3 \, c_2^2 \, \beta_{\text{inter}} 
   & -3 \, c_2^5 \, \beta_{\text{inter}}-3 \, c_2^3 \, \beta_{\text{inter}} \\ \\
 3 \, c_2^1 \,\beta_{\text{inter}}-9 \, c_2^3 \, \beta_{\text{inter}} 
 & 3 \, c_2^2 \, \beta_{\text{inter}}-9 \, c_2^4 \, \beta_{\text{inter}} 
 & 3 \, c_2^3 \, \beta_{\text{inter}}-9 \, c_2^5 \, \beta_{\text{inter}} 
   & 3 \, c_2^4 \, \beta_{\text{inter}}-9 \, c_2^6 \, \beta_{\text{inter}} \\ \\
 9 \, c_2^2 \, \beta_{\text{inter}}-3 \, c_2^0 \, \beta_{\text{inter}} 
 & 9 \, c_2^3 \, \beta_{\text{inter}}-3 \, c_2^1 \,\beta_{\text{inter}} 
 & 9 \, c_2^4 \, \beta_{\text{inter}}-3 \, c_2^2 \, \beta_{\text{inter}} 
   & 9 \, c_2^5 \, \beta_{\text{inter}}-3 \, c_2^3 \, \beta_{\text{inter}} \\ \\
 15 \, c_2^3 \, \beta_{\text{inter}}-9 \, c_2^1 \,\beta_{\text{inter}} 
 & 15 \, c_2^4 \, \beta_{\text{inter}}-9 \, c_2^2 \, \beta_{\text{inter}} 
 & 15 \, c_2^5 \, \beta_{\text{inter}}-9 \, c_2^3 \, \beta_{\text{inter}} 
   & 15 \, c_2^6 \, \beta_{\text{inter}}-9 \, c_2^4 \, \beta_{\text{inter}} \\ \\
 1 + 3 \, c_2^2 \, \beta_{\text{intra}}^{3/2, 3/2}-5 \, c_2^0 \, \beta_{\text{intra}}^{3/2, 3/2}
 & 3 \, c_2^3 \, \beta_{\text{intra}}^{3/2, 3/2}-5 \, c_2^1 \,\beta_{\text{intra}}^{3/2, 3/2} 
   & 3 \, c_2^4 \, \beta_{\text{intra}}^{3/2, 3/2}-5 \, c_2^2 \, \beta_{\text{intra}}^{3/2, 3/2} 
   & 3 \, c_2^5 \, \beta_{\text{intra}}^{3/2, 3/2}-5 \, c_2^3 \, \beta_{\text{intra}}^{3/2, 3/2} \\ \\
 -2 \, c_2^3 \, \beta_{\text{intra}}^{3/2, 3/2}-6 \, c_2^1 \,\beta_{\text{intra}}^{3/2, 3/2} 
 & 1-2 \, c_2^4 \, \beta_{\text{intra}}^{3/2, 3/2}-6 \, c_2^2  \beta_{\text{intra}}^{3/2, 3/2}
   & -2 \, c_2^5 \, \beta_{\text{intra}}^{3/2, 3/2}-6 \, c_2^3 \, \beta_{\text{intra}}^{3/2, 3/2} 
   & -2 \, c_2^6 \, \beta
   _{\text{intra}}^{3/2, 3/2}-6 \, c_2^4 \, \beta_{\text{intra}}^{3/2, 3/2} \\ \\
 3 \, c_2^0 \, \,\beta_{\text{intra}}^{3/2, 3/2}-9 \, c_2^2 \, \beta_{\text{intra}}^{3/2, 3/2} 
 & 3 \, c_2^1\, \beta_{\text{intra}}^{3/2, 3/2}-9 \, c_2^3 \, \beta_{\text{intra}}^{3/2, 3/2} 
 & 1-9 \, c_2^4 \, \beta_{\text{intra}}^{3/2, 3/2}+3 \, c_2^2 \, \beta_{\text{intra}}^{3/2, 3/2}
 & 3 \, c_2^3 \, \beta_{\text{intra}}^{3/2, 3/2}-9 \, c_2^5 \, \beta_{\text{intra}}^{3/2, 3/2} \\\\
  0 & 0 & 0 & 1 
\end{matrix} 
\right ]
\end{align*}
}
and
{\tiny
\begin{align*}
  \begin{bmatrix}
3 \, \text{hc}_2^2 \, \beta_{\text{inter}}
+ 3  \, \text{hc}_2^0 \,  \beta_{\text{inter}}
-3  \, \text{hc}_1^2\,  \beta_{\text{intra}}^{1/2, 1/2}
+5  \, \text{hc}_1^0 \,  \beta_{\text{intra}}^{1/2, 1/2} \\ \\
 9  \, \text{hc}_2^3 \,\beta_{\text{inter}}-3  \, \text{hc}_2^{1} \, \beta_{\text{inter}}
 -27  \, \text{hc}_1^3 \,\beta_{\text{intra}}^{1/2, 1/2}+
   17  \, \text{hc}_1^{1} \, \beta_{\text{intra}}^{1/2, 1/2} \\ \\
 -9  \, \text{hc}_2^2 \,\beta_{\text{inter}}+3  \, \text{hc}_2^0 \,  \beta_{\text{inter}}
 +9  \, \text{hc}_1^2 \,\beta_{\text{intra}}^{1/2, 1/2}
 -3 \, \text{hc}_1^0 \,  \beta_{\text{intra}}^{1/2, 1/2} \\ \\
 -15  \, \text{hc}_2^3 \,\beta_{\text{inter}}
 + 9  \, \text{hc}_2^{1} \, \beta_{\text{inter}}
 +45  \, \text{hc}_1^3 \, \beta_{\text{intra}}^{1/2, 1/2}
 -27  \, \text{hc}_1^{1} \, \beta_{\text{intra}}^{1/2, 1/2} \\ \\
 3  \, \text{hc}_1^2\, \beta_{\text{inter}}+3  \, \text{hc}_1^0 \,  \beta_{\text{inter}}
 -3  \, \text{hc}_2^2\, \beta_{\text{intra}}^{3/2, 3/2}
 +5 \, \text{hc}_2^0 \,  \beta_{\text{intra}}^{3/2, 3/2} \\ \\
 9  \, \text{hc}_1^3 \, \beta_{\text{inter}}
 -3  \, \text{hc}_1^{1} \, \beta_{\text{inter}}
 +2  \, \text{hc}_2^3 \, \beta_{\text{intra}}^{3/2, 3/2}
   +6  \, \text{hc}_2^{1} \, \beta_{\text{intra}}^{3/2, 3/2} \\ \\
 -9  \, \text{hc}_1^2\, \beta_{\text{inter}}+3  \, \text{hc}_1^0 \,  \beta_{\text{inter}}
 +9  \, \text{hc}_2^2 \,\beta_{\text{intra}}^{3/2, 3/2}-3
    \, \text{hc}_2^0 \,  \beta_{\text{intra}}^{3/2, 3/2} \\ \\
 9  \, \text{hc}_1^{1} \, \beta_{\text{inter}}-15  \, \text{hc}_1^3 \,\beta_{\text{inter}}   
  \end{bmatrix},
\end{align*}
}
respectively.
The various notations used above represent the following integrals:
\begin{align}
& c_{\alpha_s}^n = \int d\theta \, {\mathcal F}_s (k,\theta)  \,\cos^n \theta  \,, \quad
\text{hc}_{\alpha_s}^n = \int d\theta \, {\mathcal F}_s (k,\theta)  \,\cos^n \theta \,, \nn &
\lbrace \alpha_{1/2}, \, \alpha_{3/2}  \rbrace = \lbrace 1, \, 2  \rbrace\,,
\quad {\mathcal F}_s (\mu,\theta) = \tau_s (\mu,\theta)\,
 \frac{\sin \theta \,k^3 
\, {\mathcal D}^{-1}_s (k, \theta)
}
{  |\boldsymbol k  \cdot {\boldsymbol{w}}_s (\boldsymbol k)  | }
\Bigg \vert_{ k  = k_F^s (\theta) } \,.
\end{align}

 
\bibliography{ref_rsw}


\end{document}